\documentclass{ws-rv9x6}
\usepackage{ws-rv-van}             
\newcommand{\be}{\begin{equation}}
\newcommand{\ee}{\end{equation}}
\newcommand{\bea}{\begin{eqnarray}}
\newcommand{\eea}{\end{eqnarray}}
\makeindex
\begin{document}
\def\pmb#1{\setbox0=\hbox{#1}
 \kern.05em\copy0\kern-\wd0 \kern-.025em\raise.0433em\box0 }

\setcounter{chapter}{1}
\chapter*{Frozen ghosts in thermal gauge field theory}{\footnotetext[1]{Contribution
to ``Fundamental Interactions---A Memorial Volume for Wolfgang Kummer'',
D.\ Grumiller, A.\ Rebhan, D.V.\ Vassilevich (eds.)}}

\author{P V Landshoff}
\address{Department of Applied Mathematics and Theoretical Physics\\
University of Cambridge, Cambridge CB3 0WA\\
pvl@damtp.cam.ac.uk}
\author[P V Landshoff and A Rebhan]{A Rebhan}
\address{Institut f\"ur Theoretische Physik, Technische Universit\"at Wien\\ 
A-1040 Wien, Austria \\ rebhana@tph.tuwien.ac.at}

\begin{abstract}
We review an alternative formulation of gauge field theories at finite
temperature where unphysical degrees of freedom of gauge fields
and the Faddeev-Popov ghosts are kept at zero temperature.
\end{abstract}

\body

\section{Introduction}

Thermal gauge field theory is a combination of two difficult areas of
physics, and so it is no surprise that some of its aspects are subtle.
An apparently simple and rather powerful formalism 
is in common use\cite{kapusta,lebellac},
which introduces propagators, a version of Wick's theorem, and even what
appear to be quantum states, all of which are thermal generalisations of their
zero-temperature counterparts. While many things may be calculated from
this formalism, sometimes it does not apply\cite{taylor}, 
and sometimes it is needlessly complicated\cite{rebhan}.
It can be valuable, therefore, to go back to first principles,
rather than making use of the conventional formalism without thinking.
Doing so, we shall find that it in the real-time formalism
it is possible and often advantageous to keep
the tree-level propagators of ghosts and
unphysical degrees of freedom free from thermal modifications
in arbitrary linear gauges including covariant gauges, 
whereas in the standard formulation
this is the case only for noncovariant gauges without propagating ghosts,
such as the axial gauge\cite{kummer}.

Already at zero temperature, there are two approaches to deriving the
gauge-field-theory formalism. The first works with field operators
and commutation relations, and introduces a space of kets. Some of these
kets do not correspond to physical states, because the fields have
unphysical degrees of freedom. It is necessary, therefore, to identify
a subset of the kets corresponding to the physical states, most simply 
those that contain no scalar or longitudinal gauge particles. However,
as was first noticed by Feynman\cite{feynman}, unless one introduces
additional ghost fields and the resulting kets, the probability of scattering
from a physical state to an unphysical one is not zero. That is, the
ghosts are needed to ensure that the $S$ matrix is unitary within the
subspace of physical states.

The other approach, which uses path integrals, does not explicitly
consider states and the ghosts have to be introduced for an apparently
very different reason. One can show that the two approaches are 
equivalent of course, but to do so is not simple: one
has to introduce the BRS operator\cite{weinberg,Hata:1980yr}. The operator approach is
closer to the physics and so it is the one we use here.

\def\h{\hfill\break}
At nonzero temperature, the propagators acquire a thermal part that
has to be added to the zero-temperature Feynman propagator. As we
shall discuss, there are two formalisms:
\begin{itemize}
\item All components of the gauge field, and the ghosts, become heated to
the temperature $T$.
\item Only the two physical degrees of freedom of the gauge field (the
transverse polarisations) acquire the additional thermal propagator;
the other components of the gauge field, and the ghosts, remain frozen
at zero temperature.  (This is for the bare propagators; self-energy
insertions in the unphysical bare propagators do depend on the
temperature.)
\end{itemize}

The second of these is the less commonly used, but in practice it is
sometimes much
simpler to apply. We shall describe it here. But before that, we go
back to basics and remind ourselves of just what thermal field theory
is trying to achieve.

\section{Basics of equilibrium thermal field theory}

For definiteness, consider QCD in Feynman gauge, though the discussion
of any gauge theory in any covariant gauge will be similar. A system
in thermal equilibrium is not in any particular quantum state; all one
knows is the probability of it being in any one of a complete set
of physical states. That is, one describes the system through a density
matrix that expresses the knowledge that it is in thermal equilibrium:
\be
\rho = Z^{-1} {\Bbb P} \exp(-H/T)
\label{density}
\ee
Here the units are such that Boltzmann's constant $k_B=1$, $H$ is the
Heisenberg-picture Hamiltonian. ${\Bbb P}$ is a projection operator
onto a complete set of physical states; we may choose to express it
in terms of a complete orthonormal set of asymptotic in-states:
\be
{\Bbb P}=\sum _i |i~\hbox{in}\rangle\langle i~\hbox{in}|
\label{projection}
\ee
$Z$ is called the grand partition function and is defined 
so as to make $\rho$ have unit trace:
\be
Z=\hbox{tr}~ {\Bbb P} \exp(-H/T)
\label{partition}
\ee
A trace is invariant under a change of the basis of states used to
calculate it: any complete orthonormal set of states may be used
and it may or may not include unphysical states, because their contribution
is removed by ${\Bbb P}$.

\section{Freezing unphysical degrees of freedom in the
real-time formalism}

Thermal field theory with gauge fields is more complicated than for
scalar fields largely because of the presence of ${\Bbb P}$. The
theory for scalar fields relies for its comparative simplicity on
the commutativity of traces, tr $AB=$ tr $BA$, but it is usually
not true that tr ${\Bbb P}AB=$ tr ${\Bbb P}BA$.

Note that the states $|i~\hbox{in}\rangle$ are ordinary zero-temperature
states, and the fields used to construct $H$ are ordinary zero-temperature
operators. The temperature $T$ comes in only in that it weights the way
that the states are combined together to construct the density matrix:
from (\ref{density}) and (\ref{projection}) 
\be
\rho= Z^{-1} \sum _i |i~\hbox{in}\rangle\langle i~\hbox{in}| \exp(-H/T)
\ee

Because we have chosen to express $\rho$ in terms of asymptotic in states,
to pick out a complete set of physical states we need to
consider only noninteracting fields.

In the real-time formalism one can switch off interactions adiabatically
at $t_i\to-\infty$, so that nonabelian gauge theories reduce to
(a number of) noninteracting abelian ones. 
Here $t_i$ refers to the point in time where the interaction-picture
operators of perturbation theory coincides with the full Heisenberg operators.

One can then reduce the interaction-picture
gauge fields to physical ones by projecting to transverse modes according to
\be
A_{\rm phys.}^{\mu}(k) = T^{\mu\nu}(k) A_\nu(k)
\ee
with
\be
T^{0\mu}=0,\quad T^{ij}=-(\delta^{ij}-{k^ik^j\over\mathbf k^2}).
\ee
The unphysical fields are given by
\be
A_{\rm unphys.}^\mu(k) = \left(g^{\mu\nu}-T^{\mu\nu}(k)\right) A_\nu(k),
\ee
and the ghost fields $\bar c$, $c$.

With this decomposition the corresponding parts in the free Hamiltonians
commute and one can factorise
\begin{eqnarray}
&&\sum_{i}\langle i~\hbox{in}|e^{-\beta H_{0I}} \cdots A_{\rm phys.}  \cdots
A_{\rm unphys.} \cdots \bar c \cdots c \cdots |i~\hbox{in}\rangle \nonumber\\
&=&\sum_{i}\langle i~\hbox{in}|e^{-\beta H_{0I}^{\rm phys.}}
\cdots A_{\rm phys.}  \cdots|i~\hbox{in}\rangle\nonumber\\
&&\qquad \times \langle 0|\cdots
A_{\rm unphys.} \cdots \bar c \cdots c \cdots |0\rangle
\end{eqnarray}
where $H_{0I}$ is the free Hamiltonian in the interaction picture
and $|i~\hbox{in}\rangle$ are states obtained by acting exclusively with
operators for the physical fields onto the vacuum state.\footnote{%
It is of course still true that there is a many-one correspondence 
between physical states and the kets that represent them. There is also still
the issue of indefinite metrics and negative-norm states.
The probability of scattering into any given unphysical state is in fact
not zero,
but it is cancelled by the probability of scattering into other
unphysical states.}
This leads to a perturbation theory where only 
the propagator for the physical gauge field $A_{\rm phys.}^{\mu}$
is thermal and all other propagators remain
as at zero temperature.

In the real-time formalism, propagators have a 2$\times$2 matrix structure, which
(with the Schwinger-Keldysh choice of complex time path) reads
\be
iD^{\mu\nu}(x)=\begin{pmatrix}
\langle \mathrm T A^\mu(x) A^\nu(0) \rangle & \langle A^\mu(0) A^\nu(x) \rangle \\
\langle A^\mu(x) A^\nu(0) \rangle & 
\langle \tilde{\mathrm T} A^\mu(x) A^\nu(0) \rangle 
\end{pmatrix}
\ee
where T and $\tilde{\mathrm T}$ refer to time and anti-time ordering, respectively.
In particular, a massless scalar (momentum-space) propagator reads
\be
iD=M
\begin{pmatrix}
\frac{i}{k^2-i\epsilon} & 0 \\ 0 & \frac{-i}{k^2+i\epsilon}
\end{pmatrix}
M
\ee
with
\be
M=\sqrt{n(|k_0|)}\begin{pmatrix}
e^{\beta|k_0|/2} & e^{-\beta k_0/2} \\
e^{\beta k_0/2} & e^{\beta|k_0|/2} 
\end{pmatrix},
\ee
where $n$ is the Bose-Einstein distribution function.
When only transverse gauge field modes have a nontrivial density matrix,
this matrix structure applies only to the $T^{\mu\nu}$ projection of the
gauge field propagator, whereas its complement is to be taken at zero
temperature. So the latter as well as the
ghost propagator involves the zero-temperature limit of the
matrix $M$,
\be
M_0=\begin{pmatrix}
1 & \theta(-k_0) \\ \theta(k_0) & 1
\end{pmatrix}.
\ee
In Feynman gauge, the gauge field propagator thus reads
\be
D^{\mu\nu}=-T^{\mu\nu}D-(g^{\mu\nu}-T^{\mu\nu})D_0,
\ee
which can be easily generalized\cite{rebhan2} to arbitrary linear gauges by replacing
$g^{\mu\nu}$ in the above expression by the corresponding Lorentz structure
appearing in the zero-temperature propagator.

Using these propagators, one can rather easily verify explicitly the gauge fixing
independence of hard thermal loops\cite{Frenkel:1989br,Braaten:1989mz}
and also of the thermodynamic potential at the multi-loop level.\cite{rebhan}

However, a subtlety appears in applications of the hard-thermal-loop resummation
program. Upon resummation of hard thermal loops, the gauge field propagator
has not only physical poles corresponding to transverse polarizations, but
also a collective mode with spatially longitudinal polarization\cite{Weldon:1982aq}. One can show\cite{rebhan2} that after resumming the hard-thermal-loop self-energy, the spatially longitudinal
propagator component acquires the usual matrix structure of a propagator
at finite temperature, which is in fact necessary so that no pinch singularities
appear at higher orders of the loop expansion.

\bibliographystyle{ws-rv-van}
\bibliography{ws-rv-sample}

\begin{thebibliography}{10}
\bibitem{kapusta} J Kapusta, {\it Finite-temperature field theory}, Cambridge University Press
(1989)
\bibitem{lebellac} M Le Bellac, {\it Thermal field theory}, Cambridge University Press (1996)
\bibitem{taylor} P V Landshoff and J C Taylor, 
  {\it Nuclear Physics  B} {\bf 430} (1994) 683
\bibitem{rebhan} P V Landshoff and A Rebhan, {\it Nuclear Physics B} 
{\bf 383} (1992) 607
\bibitem{kummer} W Kummer, {\it Acta Physica Austriaca} {\bf 14} (1961) 149; {\it ibid.} {\bf 41} (1975) 315
\bibitem{feynman} R P Feynman, {\it Acta Physica Polonica} {\bf 24} (1963) 697
\bibitem{weinberg} S Weinberg, {\it The quantum theory of fields}, volume II, Cambridge University Press (1996)
\bibitem{Hata:1980yr}
  H.~Hata and T.~Kugo,
  {\it Physical Review  D} {\bf 21} (1980) 3333
\bibitem{Frenkel:1989br}
  J~Frenkel and J~C~Taylor,
  {\it Nuclear Physics  B} {\bf 334} (1990) 199
\bibitem{Braaten:1989mz}
  E~Braaten and R~D~Pisarski,
  {\it Nuclear Physics  B} {\bf 337} (1990) 569
\bibitem{rebhan2} P V Landshoff and A Rebhan, 
{\it Nuclear Physics B} {\bf 410} (1993) 23
\bibitem{Weldon:1982aq}
  H~A~Weldon,
  {\it Physical Review  D} {\bf 26} (1982) 1394


\end{thebibliography}
\printindex

\end{document}